\documentclass[12pt]{article}
\usepackage{graphicx}
\usepackage{lscape}
\usepackage{epsfig}
\usepackage{citesort}
\usepackage{amssymb}
\usepackage{amsmath}
\def\ar{ \!\!\!}

\newcommand{\be}{\begin{equation}}
\newcommand{\ee}{\end{equation}}
\newcommand{\bea}{\begin{eqnarray}}
\newcommand{\eea}{\end{eqnarray}}

\def\R1{\varepsilon_1}
\def\E8{\varepsilon_8}

\def\s1{\hat s}

\def\beq{\begin{equation}}
\def\eeq{\end{equation}}
\def\bea{\begin{eqnarray}}
\def\eea{\end{eqnarray}}
\def\beeq{\begin{eqnarray}}
\def\eeeq{\end{eqnarray}}

\def\ba{\begin{array}}
\def\ea{\end{array}}

\begin{document}

\title{ {\Large \textbf{Electromagnetic Origin for Planck Mass and Dark Energy} } }
\author{\vspace{1cm} \\
{\small M.Ar{\i}k$^1$\thanks{%
e-mail:metin.arik@boun.edu.tr}\,\,, N. Kat{\i}rc{\i}$^{1,2}$\thanks{%
e-mail: nkatirci@dogus.edu.tr}} \\
{$^1$ \small Department of Physics,
Bo\u gazi\c ci University, Bebek Istanbul, Turkey }\\
{$^2$ \small Department of Physics,
 Do\u gu\c s
University}, {\small Ac{\i}badem Istanbul, Turkey }\\
       }
\date{}

\begin{titlepage}
\maketitle
\thispagestyle{empty}
\begin{abstract}
The origin of dark energy remains to be one of the challenges of modern cosmology.
We modify Jordan-Brans-Dicke theory using a vector field instead
of a scalar field and theory becomes similar to a simple
Einstein-aether theory. The time component of the vector field
picks up a cosmological background value. Identifying the vector
field to be the photon field, a small photon mass leads to late
time inflation. The time dependent background electrical potential
of the photon permeates the universe and explains the weakness of
the gravitational interaction by coupling to curvature. This
theory relates the smallness of the photon mass to the smallness
of the Hubble parameter. The model predicted photon mass is far
below observational constraints.
\end{abstract}
~~~PACS number(s): 98.80.-k, 04.50.Kd, 98.80.Es, 13.40.-f,
03.50.De, 41.20.-q
\end{titlepage}

\section{Introduction}

The universe is expanding at an ever increasing rate
 according to observations of supernovae (SNe). This
has been known since $1998$
\cite{perlmutter,riess,supernova,kowalski}, for a review see
\cite{expansion}. Cosmologists have studied to explain the source
of this behavior by modifying Einstein's theory of general
relativity or proposing new types of gravitational theories.
Cosmological constant $\Lambda$ in $\Lambda$CDM, is
currently the best candidate for the source of the accelerated
expansion. It may arise from vacuum fluctuations, however there is
a large (at least sixty orders of magnitude) discrepancy between
the predicted energy density of the vacuum in particle physics and
energy density of the cosmological constant from fitting of
$\Lambda$CDM model to observations such as Supernova type $1$A
explosions, cosmic microwave background radiation (CMBR) and
baryon acoustic oscillations (BAO) \cite{kowalski,cmb,tegmark}.

Einstein-aether theories have been recently revived with
 the purpose of fixing the divergences of quantum field theory (QFT) by breaking
Lorentz invariance and putting a short distance - large energy
cutoff for energy and momentum \cite{ted1}. Minkowski spacetime is
invariant under Lorentz transformations, which are true in all
inertial frames. When gravity comes into play, Minkowski spacetime
turns into Friedmann-Robertson-Walker spacetime,
 which models expansion of space and Lorentz invariance is broken, whereas spatial isotropy
and homogeneity are preserved. Lorentz symmetry is locally a good
symmetry of space-time, however on large scales, it may be
broken by the buildings of the universe such as matter or
radiation. For a recent discussion see \cite{ted}. If the background value of the vector field chooses a preferred
spatial direction, it contradicts the isotropy of the universe,
stated by the cosmological principle. A vector field that has zero
spatial components, but nonzero time component also breaks local
Lorentz symmetry down to the rotation subgroup. Will and Nordtvedt
investigated detectable effects of this phenomenon on the motion
of the solar system relative to a preferential reference frame
\cite{will}. In \cite{mota} it was stated that perturbation
spectra do not stringently depend on Lorentz invariance breaking
model parameters, and studies on the compact stars or black holes
(BH) \cite{ctrtb} can be more effective to determine the
constraints.

The cosmological motivation for the vector field is the exact
knowledge about the presence of the vector field when compared to
that of the scalar field. The possible relation between the photon
mass and Hubble parameter comes to mind from the approximately
same experimental values. In \cite{prokopec} the Hubble parameter
in the early inflation era is calculated as $H_I \approx 10^{21}
eV$ and it is stated that photon acquires a mass ($m_{\gamma,I}
\approx 10^{21} eV$) of the same order of magnitude in that era.
This predicts that the current phase of small acceleration causes
a nonzero, but very small photon mass of the order of magnitude
$H_0=10^{-33} eV$ \cite{prokopec}.

The best current laboratory bound on the photon mass $10^{-14}$
eV, derived from measurements of potential deviations from the
Coulomb law \cite{coulomb}, is far above the cosmological
constraint on the photon mass. From the measurements of Earth's
magnetic field \cite{earth} and the Pioneer-$10$ measurements of
Jupiter's magnetic field \cite{pioneer}, $m_{\gamma}$ is obtained
approximately as $10^{-15} eV$. Whereas an upper limit of $10^{-27} eV$ has been determined using
effects of photon mass to galactic magnetic fields
\cite{vectorgalactic}. The cosmological evolution of the electric
potential has been considered in \cite{maroto} and where it is stated that the value of the electric potential during the
early universe is $10^{-3} eV$ and it evolves to $10^{27} eV$ in
the present era \cite{maroto}.

This paper is based on the idea that quantum electrodynamics (QED)
may affect physics in long scales in a different manner, when
compared to its small scale behavior. We realize a model which shows
the relation between the photon mass and the Hubble parameter.
However, our model is restricted to the present era, the mass
parameter is a constant and thus can not explain the reason for
the huge difference between the Hubble parameters (the rate of
expansion) at the primordial and the present era. We modify
Jordan-Brans-Dicke theory (JBD) in a way similar to
Einstein-aether theory (AE) using a Lagrange multiplier field to
impose the condition that the Brans-Dicke (BD) field (the square
of the Jordan scalar field) is equal to the norm of the
electromagnetic vector potential. Different cosmological applications of Lagrange multiplier
and their implications such as Lorentz symmetry breaking and power counting renormalizable gravitational theories have also been investigated in
\cite{nojiri,capozzi,gao,lim,cai,kluson,capozziiki,nojiriiki}.

The outline of the paper is as follows. In section 2, we present
the model and its cosmological solutions for otherwise empty
space-time. We also consider a matter dominated and radiation dominated energy momentum
tensors and show that it leads to the standard $\Lambda$CDM result. We discuss the physical photon mass predicted by our model. The last section encompasses our concluding remarks.

\section{The Model}
The proposed action is
\begin{eqnarray}
S= \int d^{4}x\sqrt{-g}[- \frac{\varphi^2}{8\omega}R+\frac{1}{2}
\nabla _{\alpha }\varphi
\nabla^{\alpha}\varphi-\frac{\lambda^2}{2}(\varphi^2-A_{\alpha}A^{\alpha})-\frac{1}{4}F^{\mu\nu}F_{\mu\nu}-\frac{1}{2}m^2
A_{\alpha}A^{\alpha}]+S_M \nonumber \\
\label{action}
\end{eqnarray}%
where $S_M$ is the matter action. $\varphi$ is the Jordan field, $\lambda$ is the Lagrange multiplier field which imposes $\varphi^2=A^2$.
Equations of motion, obtained from the variation with respect to
metric, $A_{\mu}$, $\varphi$ and $\lambda$ are given as,
\begin{eqnarray}
\ar~~\frac{1}{\sqrt{-g}}\frac{\delta S}{\delta g^{\mu\nu}}=
-\frac{\varphi^2}{8\omega}G_{\mu\nu}-\frac{1}{8\omega}(\square
{\varphi^2}g_{\mu\nu}-\nabla_{\mu}\nabla_{\nu}\varphi^2)
\ar~~+\frac{1}{2}\partial _{\mu }\varphi\partial _{\nu
}\varphi-\frac{1}{4}g_{\mu\nu}g^{\alpha\beta}\partial _{\alpha
}\varphi
\partial _{\beta }\varphi \nonumber \\
\ar+\frac{\lambda^2}{2} A_{\mu}A_{\nu}+\frac{1}{4}g_{\mu\nu}\lambda^2(\varphi^2-g^{\alpha\beta} A_{\alpha}A_{\beta})
\ar~~ +\frac{1}{4}g_{\mu\nu}m^2 A_{\alpha}A^{\alpha}-\frac{1}{2}m^2
A_{\mu}A_{\nu}  \nonumber \\
\ar -\frac{1}{2}(F_{\mu\beta}F_{\nu}^{\beta}-\frac{1}{4}g_{\mu\nu}F_{\alpha
\beta}F^{\alpha\beta})+\frac{T_{\mu\nu}}{2}=0  \nonumber \\
\label{g1}
\end{eqnarray}%

\begin{eqnarray}
\ar~~\frac{1}{\sqrt{-g}}\frac{\delta S}{\delta
A_{\mu}}=\nabla_{\nu}F^{\mu\nu}+\lambda^2 A^{\mu}-m^2
A^{\mu}&=&0 \label{g2}
 \end{eqnarray}%
\begin{eqnarray}
\ar~~  \frac{1}{\sqrt{-g}}\frac{\delta S}{\delta \varphi}=
-\frac{\varphi}{4\omega}R-\square{\varphi}-\lambda^2\varphi&=&0
\label{g3}
 \end{eqnarray}%
\begin{eqnarray}
\ar~~  \frac{1}{\sqrt{-g}}\frac{\delta S}{\delta \lambda}=\lambda(g^{\alpha\beta}A_{\alpha}A_{\beta}-\varphi^2)=0.
\label{g4}
 \end{eqnarray}%
 We use a metric signature $(+---)$ for Friedmann-Robertson-Walker (FRW)
metric with flat space-like sections. The non-zero components of
the Ricci tensor and the
Ricci scalar are given by $R_{00}=-3\ddot{a}/a$, $R_{\alpha \beta }=\left( a%
\ddot{a}+2\dot{a}^{2}\right) \delta _{\alpha \beta }$, $\alpha,\beta =1,2,3$ and $R=-6\left( \ddot{%
a}/a+\dot{a}^{2}/a^{2}\right) $, respectively. $T_{\mu\nu}$ is the
energy-stress tensor for a perfect fluid given by $T_{\nu }^{\mu
}=diag\left( \rho,-p,-p,-p\right).$ We consider a cosmological
background value as
 $<A_\mu>=(A_0(t),0,0,0)$ which by \eqref{g4} gives that the scalar field $\varphi(t)$ equals $A_0(t)$ for nonzero $\lambda$.
For $\lambda=0$ and $A_{\mu}=0$, standard Jordan-Brans-Dicke
theory is obtained \cite{bransdicke}. Using equation \eqref{g2},
$\lambda^2=m^2$ and we obtain Eqs. \eqref{rho}-\eqref{phi}. The
fractional rate of change of the scale factor, Hubble
parameter($H$), and a constant fractional rate of change for the
scalar (or time component of the vector field ) field, B are
defined as, respectively
\begin{eqnarray}
 H=\frac{{\dot{a}(t)}}{a(t)} ~~~,~~ B=\frac{\dot{\varphi}(t)}{\varphi(t)}=\frac{\dot{A_0}(t)}{A_0(t)},
\end{eqnarray}
where $a$ is the scale factor and field equations above are
written in terms of H, B become
\begin{eqnarray}
 6H B+3H^2-2\omega B^2-2\omega m^2=\frac{4\omega}{\varphi^2}\rho_M \label{rho},
\end{eqnarray}%
\begin{eqnarray}
-2\dot{B}-B^2(4+2\omega)-4H B-2\dot{H}-3H^2+2\omega
m^2=\frac{4\omega}{\varphi^2} p_M, \label{pres}
\end{eqnarray}%
\begin{eqnarray}
\dot{B}+B^2+3H B=\frac{1}{2\omega}(3\dot{H}+6H^2-2\omega m^2),
\label{phi}
\end{eqnarray}%
$\rho_M$ and $p_M$ are the energy density and pressure of matter.
Note that, the electromagnetic field
\begin{eqnarray}
F_{\mu\nu}=\partial_{\mu}A_{\nu}-\partial_{\nu}A_{\mu}
\label{fmunu}
\end{eqnarray}
does not contribute to the Eqs. \eqref{rho}-\eqref{phi} since for
$A_0$ being only function of time and $A_i=0$, it becomes zero.
Thus Eqs. \eqref{rho}-\eqref{phi} are similar to the equations
obtained from a massive Brans-Dicke Model without any
electromagnetic field \cite{calikarik,sheftelarik}.
\subsection{Vacuum Solution}
The vacuum solution ($\rho=p=0$) yields a solution giving a
constant Hubble parameter (H) and the rate of change of the scalar
or vector field (B) as
\begin{eqnarray}
B=\frac{H}{2(\omega+1)} ~~,~~
H^2=\frac{4\omega(1+\omega)^2}{(2\omega+3)(3\omega+4)}m^2.
\label{sol}
\end{eqnarray}
Since the limit on the BD parameter is $\omega>10^4$
\cite{wb,wi,wuc,wd}, m should be less than $10^{-35}$ eV which is
far below the observational constraints mentioned above. We identify this solution as the dark energy solution for $t \rightarrow \infty$.
Considering only the leading terms in $\omega$, H and B for the dark energy era become
\begin{eqnarray}
 H_{\infty}=\sqrt{\frac{2\omega}{3}}m ,~~~ B_{\infty}=\frac{H_{\infty}}{2\omega}.
\label{hinfinity}
\end{eqnarray}
Now, we consider that the relation
\begin{eqnarray}
B=\frac{1}{2\omega}H
\label{BHrelation}
\end{eqnarray} which is obtained from equation \eqref{sol} by neglecting higher order terms in $\frac{1}{\omega}$ and place it into equation \eqref{phi}.
Again neglecting higher order terms in $\frac{1}{\omega}$, we obtain
\begin{eqnarray}
 H=H_{\infty}\coth({\frac{3H_{\infty}t}{2}})  , ~~~ a=a_1\sinh^{\frac{2}{3}}(\frac{3H_{\infty}t}{2})
\label{hvariable}
\end{eqnarray}
which gives $p=0$ in equation \eqref{pres} and
\begin{equation}
H^2=H_{\infty}^2(1+(\frac{a_1}{a})^{3}).
 \label{matterfried}
\end{equation}%
Thus this is natural that the matter dominated era is followed by the dark energy era in this model.
This means the solution theoretically knows how should a matter dominated era be experienced on the timeline to the late time
inflationary (dark energy dominated) era.

Denoting derivatives with respect to $a$ by prime, one can integrate
\begin{equation}
\frac{\varphi^{ \prime }}{\varphi}=\frac{B}{aH}.
\label{fidotfi}
\end{equation}
Using the relation between H and B in equation \eqref{sol} yields
\begin{equation}
|\varphi ^{2}|=|\varphi _{0}^{2}|\left[ \frac{a}{a_{0}}\right] ^{\frac{1}{%
(1+\omega )}} \label{fikare}
\end{equation}%
Since $\omega > 10^4$ $\varphi$ is nearly constant, thus the
electric potential of the vector field permeates all the universe
and its value is found as approximately $10^{30}$ eV from
\begin{equation}
\frac{1}{16 \pi G_N}=\frac{M_p^2}{2}=\frac{A_0^2}{8 \omega}=\frac{\varphi_0^2}{8 \omega}
\end{equation}
similar to the result of \cite{maroto} where $A_0$ has magnitude
as $10^{27}$ eV for the late dark energy era. The mass of the
photon behaves as dark energy and the coupling of the vector field
to gravitation is interpreted as the Planck Mass. The difference
of this model is the non-minimal coupling of gravitation to the
scalar field which is identified with the norm of the vector
field. This interpretation provides to relate the smallness of the Hubble parameter that expands the universe to the smallness of the photon
mass.

Current measurements show that the present content of matter
density in the universe is $25\%$ and the dark energy content is
$75\%$ and asymptotically universe will contain only dark energy.
Our model predicts the same results as the standard model of
cosmology and interprets the dark energy as electromagnetic dark
energy.
\subsection{An Alternative Approach to the Matter Dominated Solution}
When matter is put into Eqs. \eqref{rho} and \eqref{pres} as $\rho _{M}=\rho _{0}\left( \frac{a}{%
a_{o}}\right) ^{-3}$ and $p_{M}=0$, the relation in
Eq.\eqref{sol} between H and F is preserved. Eq.\eqref{phi} is also satisfied and the Hubble parameter is derived
as
\begin{eqnarray}
H^{2}=\frac{4\omega (1+\omega )^{2}}{(3\omega +4)(2\omega +3)}
\left[ m^{2}+\frac{2\rho _{0}}{\left\vert \varphi _{0}\right\vert
^{2}}\left(\frac{a}{a_{0}}\right) ^{\alpha } \right] \label{h2mat}
\end{eqnarray}
where $\alpha =-\left( 3+\frac{1}{1+\omega}\right)$. As mentioned above, although the matter is not put into  Eqs. \eqref{rho} and \eqref{pres},
Hubble parameter in the $\omega \rightarrow \infty$ limit for vacuum case demonstrates the same behaviour as in Equation \ref{matterfried}.

\subsection{Photon Mass}
The Proca Lagrangian density for a massive photon is given by
\begin{eqnarray}
\pounds=-\frac{1}{4}F^{\mu\nu}F_{\mu\nu}+\frac{1}{2}m_{\gamma}^2A_{\mu}A^{\mu}
\label{proca}
\end{eqnarray}%
where $m_{\gamma}$ is the photon mass. To extract $m_{\gamma}$
from Eq.\eqref{action}, one needs to substitude $\phi^2=A_{\alpha}A^{\alpha}$ and then look at the coefficients of
all the terms $A_iA^i$ to find the mass of the physical photon.
Using the relation between H and m,
\begin{eqnarray}
m_{\gamma}^2=-\frac{R}{4\omega} -m^2=\frac{\omega
(6\omega+7)}{(2\omega+3)(3\omega+4)}m^2
\label{photonmass}
\end{eqnarray}%
this equation predicts a photon mass on the order of $10^{-35}
eV$ for the vacuum solution.

Note that in our Lagrangian flat space limit of the photon mass
is given by $m_{\gamma}^2=-m^2$. Without this feature, our model would
not work. Since during the dark energy era $R=const.$, the term
$A_{\mu}A^{\mu}R$ also behaves as the photon mass so that
$m_{\gamma}^2$ is positive in equation \eqref{photonmass}. This is reminiscent of
the Higgs model where the mass term has the wrong sign, which is
corrected by the $\phi^4$ interaction via spontaneous symmetry
breaking.

We show that in our model there are two contributions to the
photon mass. The first is related to the scalar curvature of the
universe whereas the other comes from the bare photon mass in the
action. In the late universe, the two contributions are comparable
and partly cancel each other.
\subsection{Radiation Dominated Solution}
When we neglect $\omega m^2$ terms then $ a\propto\sqrt{t}$,  $\varphi=const.$
 is a solution for radiation dominated era. To investigate the corrections to this solution, we expand $H$ and $B$ as power series at $t=0$.
We obtain the solution as
\bea
H=\frac{1}{2t}+H_{\infty}^2t+... , ~~~ B=\frac{2m^2}{5}t+...
\eea
which gives $a \propto \sqrt{t}e^{H_{\infty}^2t^2}$
for equation of state $p=\frac{\rho}{3}$. Note that the term $H_{\infty}^2t^2$ in the exponent
is reminiscent of inflation since it indicates a tendency for the decelerating universe $a \propto \sqrt{t}$ to accelerate. However this term is insufficient to be
interpreted as early inflation since the present lifetime of the universe of the order of $H_{\infty}^{-1}$ and during the era of radiation dominated $H_{\infty}^2t^2$ is
negligible small. We can obtain standard early inflation only by putting $p=-\rho$.

\section{Conclusion}
Cosmological background value of the time component of the photon
field, $A_{\mu}$ may explain the late time expansion of the
universe. Electric potential of the photon field causes constant Hubble parameter for the vacuum case and universe is considered as
asymtotically deSitter space. This result is consistent with seven year WMAP data. This data has been analyzed and the dark energy "equation of state" parameter is $-1.10 \pm 0.14$, consistent
with the cosmological constant (or equation of state parameter $-1$) \cite{wmapseven}. The role of the cosmological constant in our model is played by the photon mass. Another feature of
our model is that the Planck mass is interpreted as the scalar potential of the photon field that fills the universe.
The huge amount of homeogeneous electric potential may explain the weakness of the gravitational force.
The smallness of the photon mass is related to the smallness of the present Hubble parameter. The difference between the Hubble
parameter value in the primordial inflation era and present era has not yet been
explained in this model. A possible mechanism for the change of the photon
mass may be used to obtain the vastly different inflationary evolutions for both
eras. The essence of this model is the embedded timelike vector field
to Jordan-Brans-Dicke theory with the help of Lagrange multiplier
field.

\end{document}